\documentclass[namedreferences,hyperref,optionalrh,solaromanenum]{spr-sola}

\usepackage{graphicx}                    
\usepackage{amssymb,amsmath}                    
\usepackage{color}                       
\usepackage{breakurl}                         

\chardef\us=`\_

\begin{document}

\begin{frontmatter}

\title{Comparative Statistics of Solar Flares and Flare Stars}

%
\author[addressref=aff1,corref,email={katz@wuphys.wustl.edu}]{\inits{J. I. }\fnm{Jonathan }\snm{Katz}\orcid{0000-0002-6537-6514
}}
\author[addressref=aff2,email={mbharmalk@gmail.com}]{\inits{M. }\fnm{Mustafa } \snm{Bharmal}}
\author[addressref=aff3,email={matthew95129@gmail.com}]{\inits{M. }\fnm{Matthew }\snm{Ju}}
\author[addressref=aff1,email={whitsett.n@email.wustl.edu}]{\inits{N. }\fnm{Nathan }\snm{Whitsett}}
\address[id=aff1]{Department of Physics and McDonnell Center for the Space Sciences, Washington University, St. Louis, Mo. 63130 USA}
\address[id=aff2]{St. Xavier's College, Ahmedabad, India}
\address[id=aff3]{Lynbrook High School, San Jos\'e, CA 95129 USA}
%
\runningauthor{Katz {\it et al.\/}}
\runningtitle{Comparative Statistics of Solar Flares and Flare Stars}


\begin{abstract}
	The distribution of interval times between recurrent discrete
	events, such as Solar and stellar flares, reflects their underlying
	dynamics.  Log-normal functions provide good fits to the interval
	time distributions of many recurrent astronomical events.  The
	width of the fit is a dimensionless parameter that characterizes
	its underlying dynamics, in analogy to the critical exponents of
	renormalization group theory.  If the distribution of event
	strengths is a power law, as it often is over a wide range, then
	the width of the log-normal is independent of the detector
	sensitivity in that range, making it a robust metric.  Analyzing
	two catalogues of Solar flares over periods ranging from 46 days
	to 37 years, we find that the widths of log-normal fits to the
	intervals between flares are wider than those of shot noise,
	indicating memory in the underlying dynamics even over a time much
	shorter than the Solar cycle.  In contrast, the statistics of flare
	stars are consistent with shot noise (no memory).  We suggest that
	this is a consequence of the production of Solar flares in localized
	transient active regions with varying mean flare rate, but that the
	very energetic flares of flare stars result from global magnetic
	rearrangement that reinitializes their magnetohydrodynamic
	turbulence.
\end{abstract}

%

\end{frontmatter}

%
\section{Introduction}\label{intro}
It is sometimes possible to gain insight into physical phenomena without
being able to calculate them in detail.  The classic examples are the
universality classes of critical points \citep{PV02}, in which phenomena
as diverse as liquid-gas and ferromagnetic critical points may be found to
have fundamental similarities, despite their very different microscopic
physics.  In fact, these similarities may be found without quantitative
theories of the phenomena.

Many astronomical phenomena consist of separated but repeating events.
Examples include Solar flares, {the flares of} flare stars, Soft Gamma
Repeaters, repeating fast radio bursts, neutron star X-ray bursts,
recurrent nov\ae, dwarf nov\ae, pulsar glitches and pulsars themselves.
The distribution of intervals between successive events contains
information about their underlying dynamics.  {Extreme superflares on
Sun-like stars \citep{VV24} may have significant terrestrial effects if
they occur on the Sun.  This increases interest in the comparative
statistics of the giant flares of flare stars (that may be considered
superflares) and Solar flares.  These authors found that the rate of
superflares is consistent with extrapolation of observations of lesser
Solar flares, suggesting a common mechanism, although it is not possible
to compile interval statistics of such rare events.}

Some recurrent events are very accurately periodic, with the intervals
between successive events almost exactly the same, at least over feasible
durations of observation.  For example, pulsar pulses repeat with the very
stable rotation period of the neutron star that emits them.  Recurrent
nov\ae\ and neutron star X-ray bursts are quasi-periodic, as accreted matter
accumulates on (respectively) a white dwarf or neutron star until there is
enough to trigger nuclear burning.  Other phenomena are far from periodic,
and the distribution of the intervals between the events is much broader.
These include Solar flares, the flares of flare stars, Soft Gamma Repeaters
and repeating fast radio bursts.

Many single-peaked distributions are well-fit by log-normal functions,
Gaussian fits to the distribution of the logarithms of the variable, in
this case the intervals (waiting times) between successive events.  A
log-normal distribution has only three parameters: its mean value (the
peak of the distribution of the logarithms of the intervals), its width
(the standard deviation of the Gaussian fit) and a normalizing factor.  The
ability of a log-normal to fit an empirical distribution does not, itself,
give insight into the underlying physics because this functional form is
very flexible.  However, the width of the fitted log-normal is analogous to
a critical point exponent of renormalization group theory:
\begin{enumerate}
	\item It is dimensionless
	\item Disparate phenomena may be united by similar log-normal widths
		(in analogy to the universality classes of renormalization
		group theory)
	\item If the distribution of event strengths is a power-law, as is
		often the case, the width is independent of the sensitivity
		of the observing system.
\end{enumerate}
{Hence the fitted log-normal width is a robust single-parameter
description of the distribution of some quantity.  That quantity may be
the strength of discrete events, or the waiting times between them.}

\citet{K24} reviewed the use of log-normal fits to the intervals between
events in several types of repeating episodic, but aperiodic, astronomical
phenomena.  That paper's emphasis was on repeating Fast Radio Bursts, but it
also presented results for Soft Gamma Repeaters (powered by the
magnetostatic energy of ``magnetars'', hypermagnetized neutron stars), the
magnitudes of and waiting times between microglitches (small sudden
increases of rotation rate) of the Vela pulsar (believed to result from
sudden increases in coupling between the rotation of solid and superfluid
components of the neutron star, but not understood in detail) and flares of
flare stars.

The widths of the log-normal fits to the distribution of intervals may be
compared to the calculated standard deviation 0.723 of the log-normal fit to
a ``shot noise'' process, a process (like radioactive decay) in which
individual events do not influence each other \citep{K24}.  A larger width
indicates memory: the object's activity varies, with active periods in which
events are frequent and intervals are short and inactive periods in which
events are infrequent and intervals are long, spreading their distribution.
A smaller log-normal width indicates a different kind of memory, like that
of a relaxation oscillator, in which events are less likely to occur shortly
after a previous event.  The limiting case of this is a periodic phenomenon
like pulsar pulses, in which the distribution of intervals is narrowly
peaked, approaching a Dirac $\delta$-function.

The distributions of variables other than intervals may also be fit by
log-normal functions.  For example, the strength of an event may be fit;
then zero width would indicate that all events have the same strength
(``standard candles''), while a broad width would indicate a broad
distribution of strengths.  The magnitudes of microglitches of the Vela
pulsar are an example; their strengths vary, but less than the intervals
between shot noise events, while the microglitch intervals are described by
shot noise statistics \citep{K24}.  More prosaic examples include relaxation
oscillators.  However, distributions of strength are often power laws, which
are not well-fit by log-normal functions.

{The statistics of waiting times between Solar flares have long been
studied, and there is an extensive literature.  Recent studies include
\citet{S20,AJ21,AJN21}, who have found and quantified evidence of memory
from their waiting time distributions.  Our log-normal fits quantify this
information in a different manner than used in the earlier work.
\citet{KM23} have calculated the \citet{LH91} model (but not the data used
here) and fitted the distribution of waiting times between extreme events
in that model with log-normals.

The novelty of this paper is the application of log-normal fits to
distributions of intervals between Solar flares and their comparison to
log-normal fits of the distributions of intervals between outbursts of
flare stars.  We consider two Solar flare databases, the recently published
\citep{V24} Chandrayaan-2 XSM Catalogue and the long-established and very
large GOES database \citep{P23}.  These two databases classify flares
differently, Chanrayaan-2 XSM into types on the basis of their temporal
structure and GOES into classes on the basis of strength.  Comparison of the
log-normal widths of waiting times between flare types or classes in the
same catalogue and between catalogues may show which have fundamentally
similar or different physics.  In fact, we find that all the varieties of
Solar flares have similar statistics, but that these differ from the
statistics of flares on flare stars.}

In contrast to the other astronomical events whose distributions of waiting
times were studied by \cite{K24}, Solar flares are much better understood,
both phenomenologically (in association with the Solar cycle) and
theoretically.  The purpose of the log-normal method is to find (or exclude)
commonalities between superficially different phenomena by comparing their
distributions of intervals.
\section{Solar Flares: The Chandrayaan-2 XSM Catalogue}
The X-Ray Solar Monitor aboard the Chandrayaan-2 satellite in a low (about
120 km above the surface) Lunar orbit \citep{M20,M21} observed 6266 Solar
flares over a three-year period from 2019 to 2022 \citep{V24}.  This
observing period began in the Solar activity minimum between sunspot cycles
24 and 25 and continued roughly half-way to the maximum of cycle 25
anticipated for 2025.  The data may be found in \citet{SD-CH2}\footnote{The
data file contains seven type B and six type A flares at h:m:s=00:00:00 that
appear to be spurious and are ignored.}.

The Lunar orbit of Chandrayaan-2 is perpendicular to the ecliptic, as shown
in Fig.~3 of \cite{M21}.  At some times (referred to as ``dawn-dusk'') its
orbital axis points to the Sun and it views the Sun continuously.  At other
times (``noon-midnight'') its orbital axis is normal to the direction
to the Sun and it is in the Moon's shadow for nearly half its orbit (Table 2
of \citet{M20}).  At these latter times, and during much of the intermediate
periods, Lunar shadowing implies that some flares are missed (Fig. 1 of
\cite{V24} illustrates the effect on the rate of flare detections) and it is
not possible to measure flare intervals.  As a result, we limit our analysis
to the periods, 46 days long and occurring every six months, during which
the Sun is continuously observed by the satellite.  These give the
flat-topped portions of the detection rates shown in Fig.~1 of \cite{V24}.
Fig.~1 (left panel) shows the counts of flares of Types A and B in the
Chandrayaan-2 XSM catalogue.
\begin{figure}
	\centering
	\includegraphics[width=0.49\textwidth,clip=]{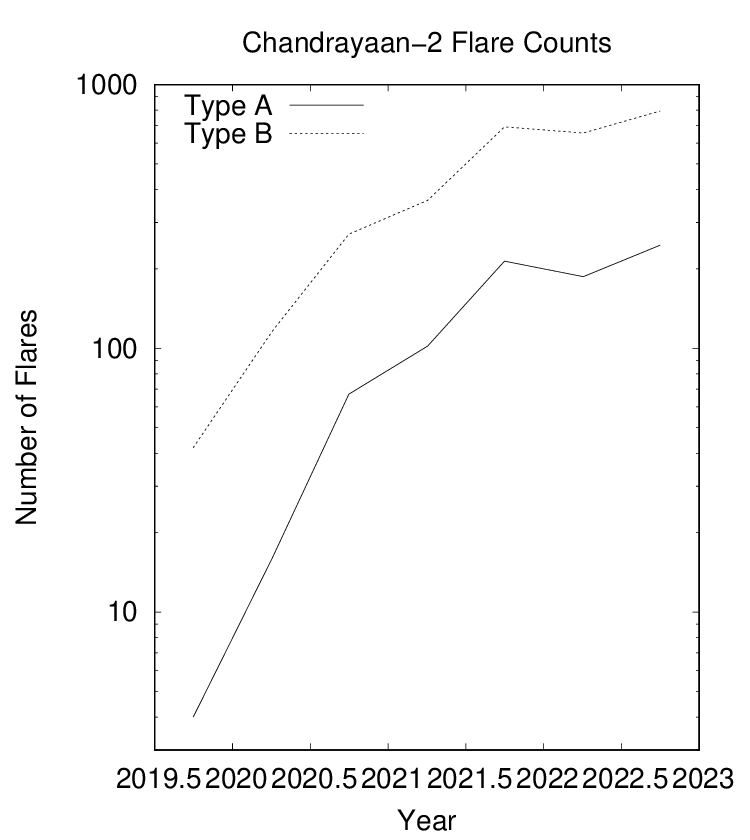}
	\includegraphics[width=0.49\textwidth,clip=]{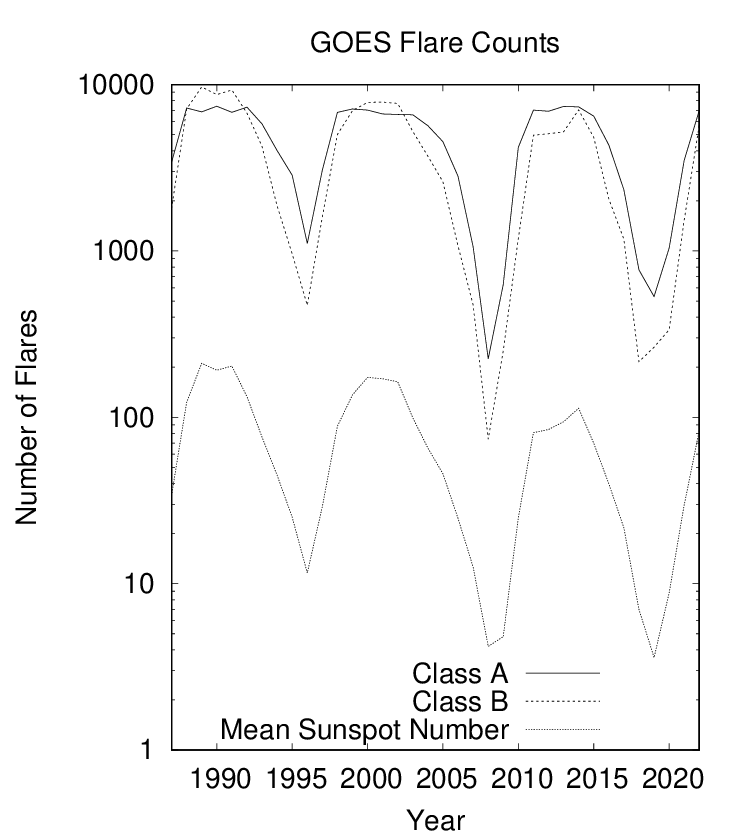}
	\caption{\label{flarerate} Left panel: {Chandrayaan-2} types A
	and B flare counts in each of the 46 day ``dawn-dusk'' periods of
	continuous Solar visibility of the Chandrayaan-2 XSM catalogue
	\citep{SD-CH2}.  Observations began during the sunspot minimum
	between Solar cycles 24 and 25 (the first five days of the September
	7--October 23, 2019 ``dawn-dusk'' period were not included), and
	continued roughly half-way to the maximum of cycle 25 predicted for
	2025.  Right panel: {Classes} A and B GOES flare counts and
	annual mean sunspot numbers 1987--2022 \citep{GOES,Sunspots}; flare
	activity follows the Solar cycle {(Chandrayaan-2 flare types are
	unrelated to GOES flare classes)}.}
\end{figure}

We fit log-normals to the distributions of intervals for flares of types A
and B (as designated in \citet{SD-CH2} {on the basis of their temporal
structure}).  Because variations in the mean flare rate broaden the
distribution of intervals, we fit log-normals to intervals in all the six 46
day ``dawn-dusk'' periods over three years taken together, during which the
mean flare rate changed by a large factor, and separately to the final 46
day ``dawn-dusk'' period of unobscured observation, during which the mean
flare rate appears to have been roughly constant (Fig.~1).

\begin{figure}
	\centering
	\includegraphics[width=0.49\columnwidth]{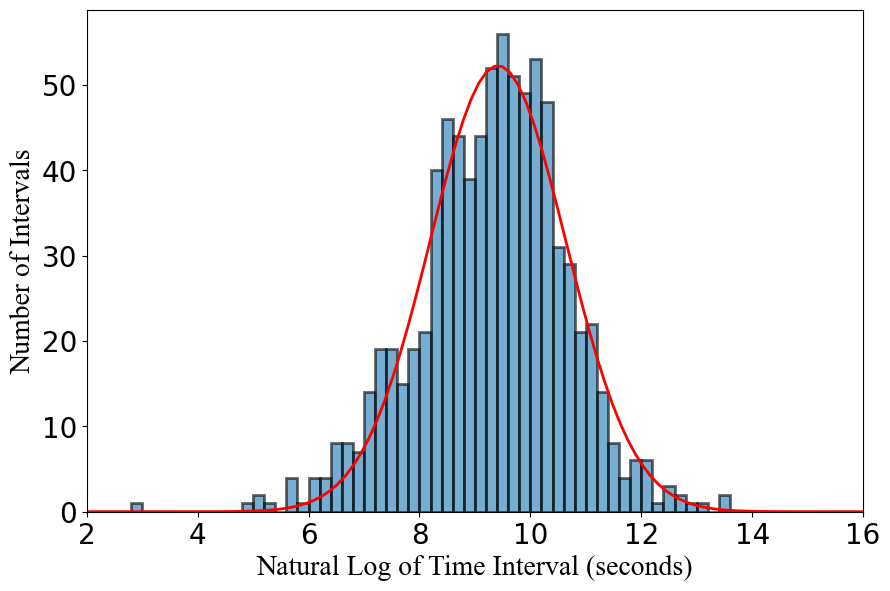}
	\includegraphics[width=0.49\columnwidth]{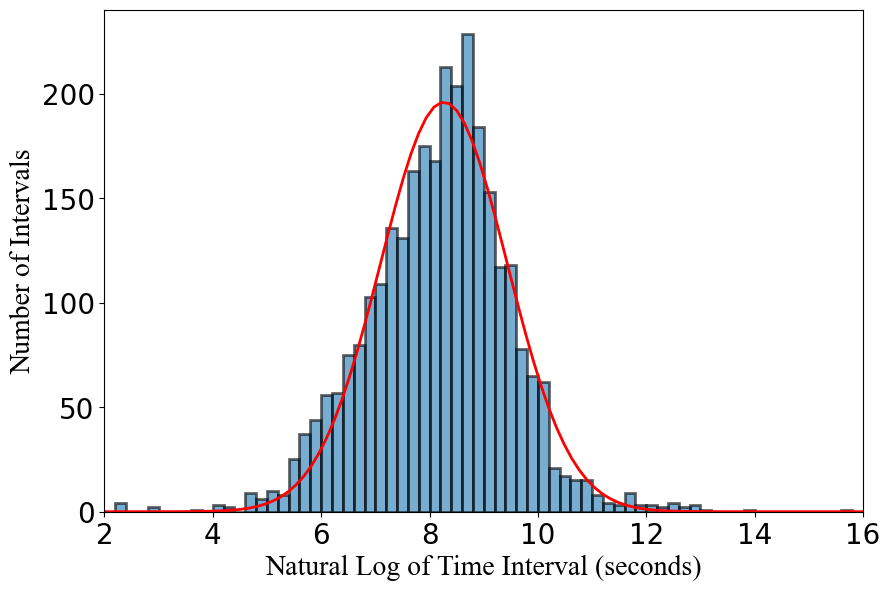}
	\caption{\label{Chandrayaan}Distributions of natural logs of
	intervals (in s) between {Chandrayaan-2} type A flares (left
	panel) and between type B flares (right panel) and log normal fits
	to these distributions for all Chandrayaan-2 data during 46 day
	periods when the Sun was never obscured by the Moon.  Solar
	activity, as measured by sunspot numbers and by flare detections,
	was rapidly increasing during this approximately 2 1/2 year period
	at the beginning of Solar cycle 25, as shown in Fig.~1.}
\end{figure}

Fig.~\ref{Chandrayaan} shows the distributions of intervals and log-normal
fits using flares from all the 46 day periods in the Chadrayaan-2 database
during which the Sun is never obscured by the Moon, taken together (the
spurious very long intervals between the last flare in one 46 day period and
the first flare in the next are ignored).  Because the mean rate of flares
varied rapidly during that time the lognormal width and kurtosis are
expected to be greater than their ``instantaneous'' values.  However, these
differences are small because the statistics are dominated by the much
larger number of intervals observed when the flare rate is higher and likely
to be roughly steady (Fig.~1 left panel).

Fig.~3 shows the distribution of intervals and log-normal fits for the final
46 day period, when sufficient flares were observed to obtain meaningful
statistics, but the mean flare rate is expected to have been approximately
steady.  Fig.~1 shows that the mean rate of flares was not rapidly
changing around that time, so that these results approximate their
``instantaneous'' values.
\begin{figure}
	\centering
	\includegraphics[width=0.49\columnwidth]{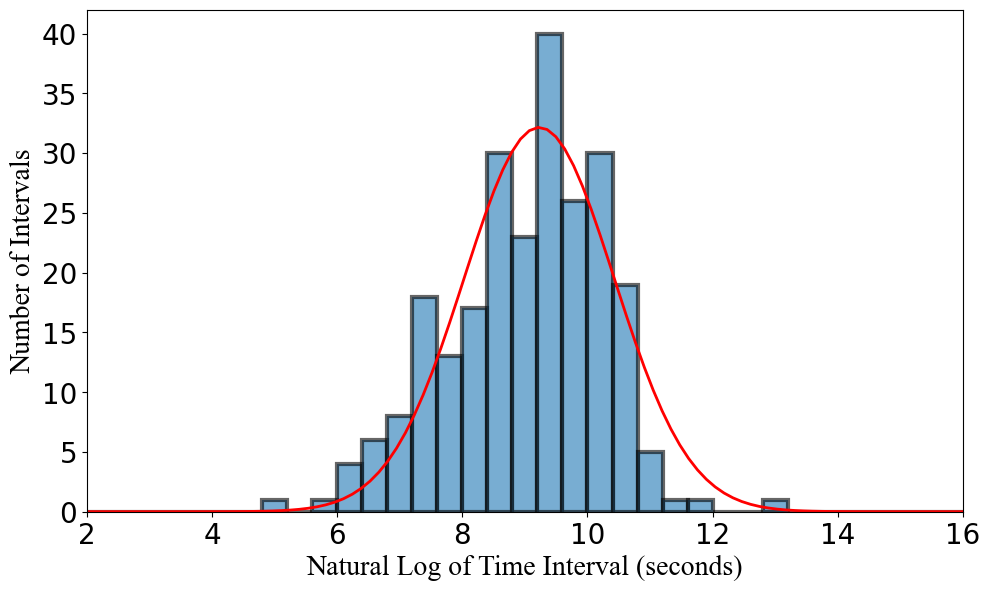}
	\includegraphics[width=0.49\columnwidth]{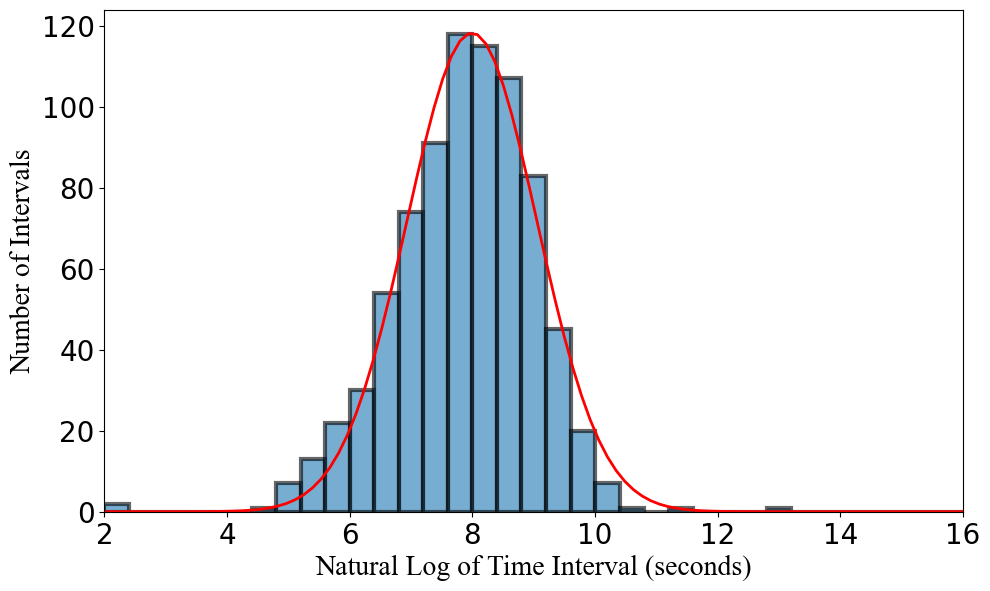}
	\caption{\label{Chandrayaanlast}Distributions of natural logs of
	intervals (in s) between Chandrayaan-2 type A flares (left panel)
	and between type B flares (right panel) and log normal fits to
	these distributions during the period September 7--October 23,
	2022 when the satellite observed the Sun continuously (without
	obscuration by the Moon).  Solar activity, as measured by sunspot
	numbers, was roughly half-way to the maximum of Solar cycle 25 and
	the mean rate of flares was not rapidly varying.}
\end{figure}

The statistics are summarized in Table~1.  The widths of the log-normals
fitted to all the (``dawn-dusk'') periods when there was no Lunar
obscuration, during which the mean flare rate changed by a large factor
(Fig.~1), are slightly, but only slightly, greater than those fitted to a
subset when the mean flare rate was not rapidly changing.  All datasets have
moderate negative skewness, indicating an excess of short intervals and a
memory effect, as expected from the variation of Solar activity as active
regions form and dissipate.  The expected excess of long intervals from
periods of low Solar activity, that would produce positive skewness, is
almost unobservable, perhaps because few flares occur during periods of low
activity.

The fact that the widths are greater than that (0.723) of shot noise, even
when the mean flare rate is not rapidly changing, implies that there are
periods of activity greater that the mean, and necessarily also periods of
lesser activity.  This is consistent with the well-known fact that the Sun
has active regions with correlated sunspot and flare activity; as these
regions grow and decay the mean flare rate, as well as the sunspot number,
increase and decrease.  This is also quantified by the excess (over 3, its
value for Gaussian statistics) kurtosis in three of the datasets,
particularly for Type B flares.
\begin{table}
	\begin{center}
	\begin{tabular}{|c|rccc|}
		\hline
		Flare Type and Period&$N$&$\sigma$&Skewness&Kurtosis\\
		\hline
		SD-CH2 Type A, Last&246&$1.22\pm 0.11$&$-0.40\pm 0.16$&$3.09\pm 0.31$\\
		SD-CH2 Type B, Last&794&$1.12\pm 0.06$&$-0.42\pm 0.09$&$4.52\pm 0.17$\\
		SD-CH2 Type A, All&1463&$1.34\pm 0.05$&$-0.32\pm 0.06$&$3.79\pm 0.13$\\
		SD-CH2 Type B, All&4790&$1.29\pm 0.02$&$-0.07\pm 0.04$&$4.38\pm 0.07$\\
		GOES 1986--2023 {B-class}&144086&$1.10\pm 0.004$&$0.61\pm 0.006$&$4.41\pm 0.013$\\
		TESS four flare stars &104--324&$0.683 \pm 0.016$&&\\
		\hline
	\end{tabular}
	\end{center}
	\caption{\label{summary}Summary of Solar flare data.  Chandrayaan-2
	XSM indicated by SD-CH2.  {Note that Chadrayaan-2 flare types are
	not the same as GOES classes, even when denoted by the same letter.}
	Uncertainties are purely statistical and do not include the effects
	of non-stationary Solar behavior.  {Variations in Solar activity
	increase $\sigma$ and kurtosis and may affect skewness in either
	direction; the separate computations for the entire Chandrayaan-2
	dataset and its last 46 day period show this effect (except for
	kurtosis).  The flare stars are TESS TOI 176.01, 218.01, 1224.01
	and 1450.01}  The uncertainty of $\sigma$ for the flare stars is the
	standard deviation of the values computed for each of the stars; the
	$N$ are too small to justify computation of skewness or kurtosis.}
\end{table}
\section{GOES Data}
The GOES satellite constellation has been collecting X-ray solar flare data
since 1986 \citep{P23}.  These satellites are in geosynchronous orbits and
are continuously illuminated by the Sun.  We analyze data \citep{GOES} from
June 1, 1986--April 30, 2023 for {Class} B flares.  {Because GOES
flares are classed on the basis of their strength, with Class B the second
weakest, the sampling of Class B flares is likely nearly complete, in
contrast to the weaker Class A flares that we ignore.}  This period includes
more than three entire solar cycles, during which the rate of Solar activity
varied greatly.  A total of 144,086 {Class} B flares were recorded
during this period; this number is larger than cited by \citet{P23} because
we use about three more years of data.  The distribution of intervals and
its fitted Gaussian is shown in Fig.~4.
\begin{figure}
	\centering
	\includegraphics[width=0.99\columnwidth]{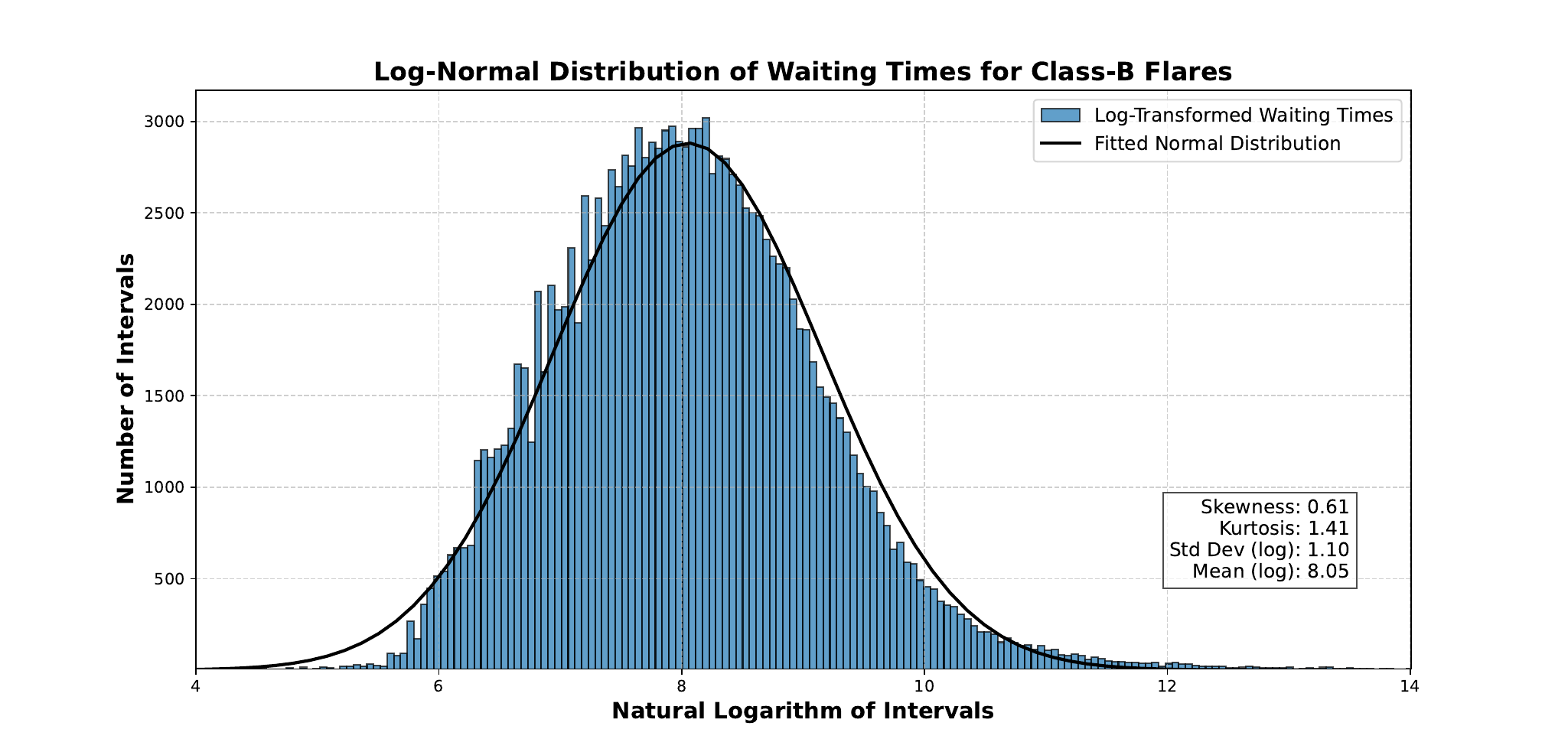}
	\caption{\label{GOES}Distributions of natural logs of intervals (in
	s) between class B flares and log-normal fit to this distribution
	for 1986--2023 GOES data.  These data span more than three Solar
	cycles.  The positive skewnesses shown in Table 1 are evident, and
	result from long intervals occurring around Solar minima.}
\end{figure}

The statistical parameters that fit the GOES data are shown in Table 1.
The large size of the database means that formal statistical uncertainties
of the fitted parameters (width, skewness and kurtosis) are negligible,
$\lesssim 0.01$.  Systematic uncertainties resulting from longer-term
variations of the Sun dominate, but cannot be estimated from shorter
datasets.

Despite the variation of the rate of Solar flare activity through the Solar
cycle, the widths of the distributions are somewhat less than those of the
Chandrayaan-2 distributions over the rising portion of one Solar cycle and
even slightly less than the Chandrayaan-2 distribution over a single period
during which the mean flare rate may be nearly constant.  Fig.~4 more
clearly shows tails of the distributions at long intervals, produced in
periods of low flare activity, that contribute to the positive skewnesses
of the GOES data.  The absence of these tails in the Chandrayaan-2 data may
possibly be attributed to its smaller dataset, so that rare long intervals
are not found at all, but may more likely be the result of greater Solar
variability in the more extended GOES dataset.  A period of 37 years (GOES)
may include a broader range of Solar states than the six 46 day periods,
distributed over about three years, of the Chandrayaan-2 data.
\section{Flare Stars}
Flare stars, dim low-mass M-dwarfs with flares that may multiply their
luminosities by large factors, are believed to resemble scaled-up versions
of Solar flares \citep{BG10}.  The TESS satellite \citep{TESS} observed a
sufficient numbers of flares from a few flare stars to permit fitting
log-normals to their distributions of interval times.

{This study used observations \citep{WD25} of four flare stars, each
with the most (between 104 and 324) TESS-detected flares, sufficient to
calculate log-normal widths with about 10\% accuracy.  Their distributions
of intervals include outlying tails of extremely short (minutes to an hour)
and extremely long (up to a year) intervals, in contrast to the peaks of
the distributions at intervals at $\sim 1\text{--}10\,$ days.  We
attribute these short intervals to substructure within flares rather than
to intervals between distinct flares (a similar excess of short intervals
between Fast Radio Bursts has long been known \citep{K18,K19}).  The
extremely long intervals are attributed to gaps in observational coverage.
We therefore exclude all intervals whose logarithms are more than five
standard deviations (of the distribution of logarithms for that particular
flare star) from the mean, iterated to self-consistency.}

The mean fitted log-normal width was $\sigma = 0.683 \pm 0.016$ (the
uncertainty is computed from the scatter of the four values of $\sigma$).
This is slightly less than the $\sigma = 0.723$ of shot noise.  The
difference has a nominal significance of 2.5 standard deviations.  {If
real, it would hint at quasi-periodicity.}  It is very significantly less
than the $\sigma = 1.1\text{--}1.3$ fitted in Table 1 to the widths of the
interval distributions of Solar flares.  The comparatively small flare star
datasets mean that the skewnesses and excess kurtoses, {more sensitive
to outliers than the widths of their interval distributions may not be
useful for comparison to these parameters for Solar flares.

However, the difference in widths between the fits to the interval
distributions of Solar flares and that of the (much more energetic) flare
stars is sufficient to establish that these phenomena differ in more than
scale.  A possible explanation may be that flares on flare stars involve a
global rearrangement of their magnetic fields (perhaps analogous to the
giant outbursts of Soft Gamma Repeaters), in contrast to the localized
nature of Solar flares.}
\section{Discussion}
The flares of flare stars have the statistics of shot noise, while Solar
flares show memory.  The latter conclusion is expected, because the Sun's
activity varies greatly with the Solar cycle (Fig.~1), broadening the
distribution of interval times.  But even in a shorter period of
Chandrayaan-2 observations during which the mean level of Solar activity is
expected to be roughly constant (Fig.~3), Solar flares show memory, as
indicated by $\sigma = 1.22 \pm 0.11$ (Chandrayaan-2 type A flares) and
$\sigma = 1.12 \pm 0.06$ (Chandrayaan-2 type B flares), differing from the
shot noise value by 4.5 and 6.7 standard deviations, respectively.
Integrating over the Solar cycle, the $\sigma$ fitted to the GOES data
differs from the shot noise value by a nominal $\sim 100$ standard
deviations.  This may be regarded only as consistency with the Solar cycle,
but comparison to the flare star data emphasizes that flare stars do not
appear to have comparable cycles, at least on the time scales of the TESS
data.

A plausible explanation is that this reflects the existence of comparatively
persistent, but not permanent, active regions on the Sun, {often only
one at a time}.  When such a region is present and facing the Earth the mean
rate of {observed} flares is higher, while when there is no such region,
or any such region is on the far side of the Sun, the mean rate of observed
flares is less.  This broadens the distribution of intervals.  In contrast,
flares on flare stars are believed (on the basis of their energy) to result
from global reorganization of the stars' magnetic fields.  Once such a
reorganization occurs, the star returns, {statistically}, to a pre-flare
state.  If the magnetohydrodynamic turbulence that builds up magnetic energy
is stochastic, it makes the distribution of intervals to the next flare
resemble that of shot noise.  This hypothesis might be explored by numerical
simulation of the turbulence, as well as by collection of additional flare
star data.

{The fact that the statistics of flare stars and Solar flares differ,
implying fundamentally different mechanisms, implies that global reorderings
of the magnetic field may not occur on the Sun.  This does not, however,
mitigate the concern \citep{VV24} that a superflare like those observed
on other Solar-type stars might occur on the Sun.}
%

%

%

%
 \begin{acks}
The author Mustafa K. Bharmal is thankful to St. Xavier's College
(Autonomous), Ahmedabad, India for providing financial assistance under
the Scheme of Research Project Grant, having award number
SXCA/XIP/2022-2023/UG05.
 \end{acks}

%
%
%
%
%
%
%

%
%
 \bibliographystyle{spr-mp-sola}
 \bibliography{solar.bib}  
%
%
%
%

\end{document}